\documentclass[12pt]{emulateapj} 

\usepackage{color}

\DeclareTextSymbol{\degre}{T1}{6}
\DeclareTextSymbol{\degre}{OT1}{23}
\usepackage{wasysym}
\usepackage{array}
\usepackage{graphicx}
\usepackage{float}
\usepackage{hyperref}
\usepackage{textcomp}
\usepackage{epstopdf}

\shorttitle{The formation of Uranus and Neptune on the CO iceline}
\shortauthors{Ali-Dib et al.}

\begin{document}


\title{The measured compositions of Uranus and Neptune from their formation on the CO iceline}

\author{Mohamad Ali-Dib,\altaffilmark{1} Olivier Mousis,\altaffilmark{1} Jean-Marc Petit,\altaffilmark{1} and Jonathan I. Lunine\altaffilmark{2}}

\email{mdib@obs-besancon.fr}

\altaffiltext{1}{Universit{\'e} de Franche-Comt{\'e}, Institut UTINAM, CNRS/INSU, UMR 6213, Observatoire de Besan\c con, BP 1615, 25010 Besan\c con Cedex, France}
\altaffiltext{2}{Center for Radiophysics and Space Research, Space Sciences Building, Cornell University, Ithaca, NY 14853, USA}


\begin{abstract}

The formation mechanisms of the ice giants Uranus and Neptune, and the origin of their elemental and isotopic compositions, have long been debated. The density of solids in the outer protosolar nebula is too low to explain their formation, and spectroscopic observations show that both planets are highly enriched in carbon, very poor in nitrogen, and the ices from which they originally formed might had deuterium-to-hydrogen ratios lower than the predicted cometary value, unexplained properties observed in no other planets. Here we show that all these properties can be explained naturally if Uranus and Neptune both formed at the carbon monoxide iceline. {Due to the diffusive redistribution of vapors,} this outer region of the protosolar nebula intrinsically has enough surface density to form both planets from carbon-rich solids but nitrogen-depleted gas, in abundances consistent with their observed values. Water rich interiors originating mostly from transformed CO ices reconcile the D/H value {of Uranus and Neptune's building blocks} with the cometary value. Finally, Our scenario generalizes a well known hypothesis that Jupiter formed on an iceline (water snowline) for the two ice giants, and might be a first step towards generalizing this mechanism for other giant planets.\\

 
 
\end{abstract}

\section{Introduction}
Uranus and Neptune are the outermost planets of the solar system. {Formation at their current positions poses the problem of how a large density of solids could have existed that far out in the protosolar nebula (hereafter PSN), since gas density is thought to decrease with the inverse heliocentric distance \citep{pol96}. A large solids surface density is needed to form the planetary cores quickly enough to accrete gas within a timescale consistent with the presence of the gaseous protoplanetary disk in the currently accepted models of giant planets formation \citep{2014arXiv1404.5018H}.} 


With atmospheric C/H ratios measured to be enhanced by factors of $\sim$30 to 60 times the solar value \citep{feg91}, both planets appear highly enriched in carbon. In comparison, the C/H ratios in Jupiter and Saturn have been measured to be about 4 and 7 times the solar value respectively \citep{gal04,fletcher2009}, and are thought to be consistent with some core-accretion formation models.\\

The nitrogen abundance is also surprising, since both planets have very low N/H ratios ($\sim$ 1\% of the solar value) \citep{dep89,dep892,gaut89}. Jupiter and Saturn on the other hand are enriched in nitrogen by a factor $\sim$ 4 compared to the solar value \citep{gal04,fletcher2009}. This large difference motivated several studies that tried to explain the N depletion in Uranus and Neptune, with little success \citep{feg91,at95}. This differential enrichment found in Uranus and Neptune, in contrast with the uniformly enriched Jupiter and Saturn, hints at differences in {the initial composition of their formation locations}.\\

The Deuterium to Hydrogen (D/H) ratio, strongly temperature dependent and considered an indicator of ices formation location, is also problematic for Uranus and Neptune. This ratio was measured in both atmospheres. These measurements were coupled to models of planets interiors \citep{helled11} to obtain the D/H ratios for the original water proto-ices that contributed in forming the planets (hereafter proto-ices). By making the assumption that the water in their interiors originated entirely from nebular H$_2$O ice, its D/H value was found $\sim$ 6-8 times lower than the cometary values {in both Oort cloud and Jupiter family comets} \citep{lis}. This is surprising because Uranus and Neptune are supposed to have formed in the region of the comets and thus their proto-ices should have cometary D/H. 

Here we show that we can explain all these unique properties at once if Uranus and Neptune formed at the CO iceline {with N$_2$ iceline a short distance outward}. In Sec. 2 we present the dynamical multi–snowline volatiles transport model we used to calculate the CO iceline properties. In section 3 we discuss the results and show how this model explain the aforementioned observations. Caveats and implications of our results are discussed in Sec. 4 and we conclude in Sec. 5.\\

\section{The volatiles distribution model}
In order to calculate the composition and properties of the CO iceline, we used the dynamical volatiles transport and distribution model from \cite{mad} (AD14). It tracks the evolution of CO and N$_2$ solids and vapor in a standard model of the PSN. This model takes into account the major dynamical and thermodynamical effects relevant to volatiles: turbulent gas drag \citep{step,hughes} and sublimation \citep{sup} for solids, in addition to gas diffusion \citep{stev88} and condensation \citep{ros} (RJ13) for vapors. We used the same modules and parameters as AD14 unless otherwise stated. The main modification is the replacement of the dust coagulation module used in AD14 with the more effective growth due the condensation from RJ13. {In this model condensation and sublimation are simulated in a Monte Carlo scheme where solid particles diffuse following turbulence leading some of them to cross an iceline and sublimate. The resulting vapor recondenses onto already existing particles, leading to {pebble growth}. For a minimum mass solar nebula,} millimetric dust are found to grow into pebbles in $10^3$ $\Omega_K^{-1}$. We used the equilibrium vapor pressure for N$_2$ from \citep{N2}. The results for each module are presented in Table 1.\\

\begin{figure*}
\begin{center}
\includegraphics[scale=0.3]{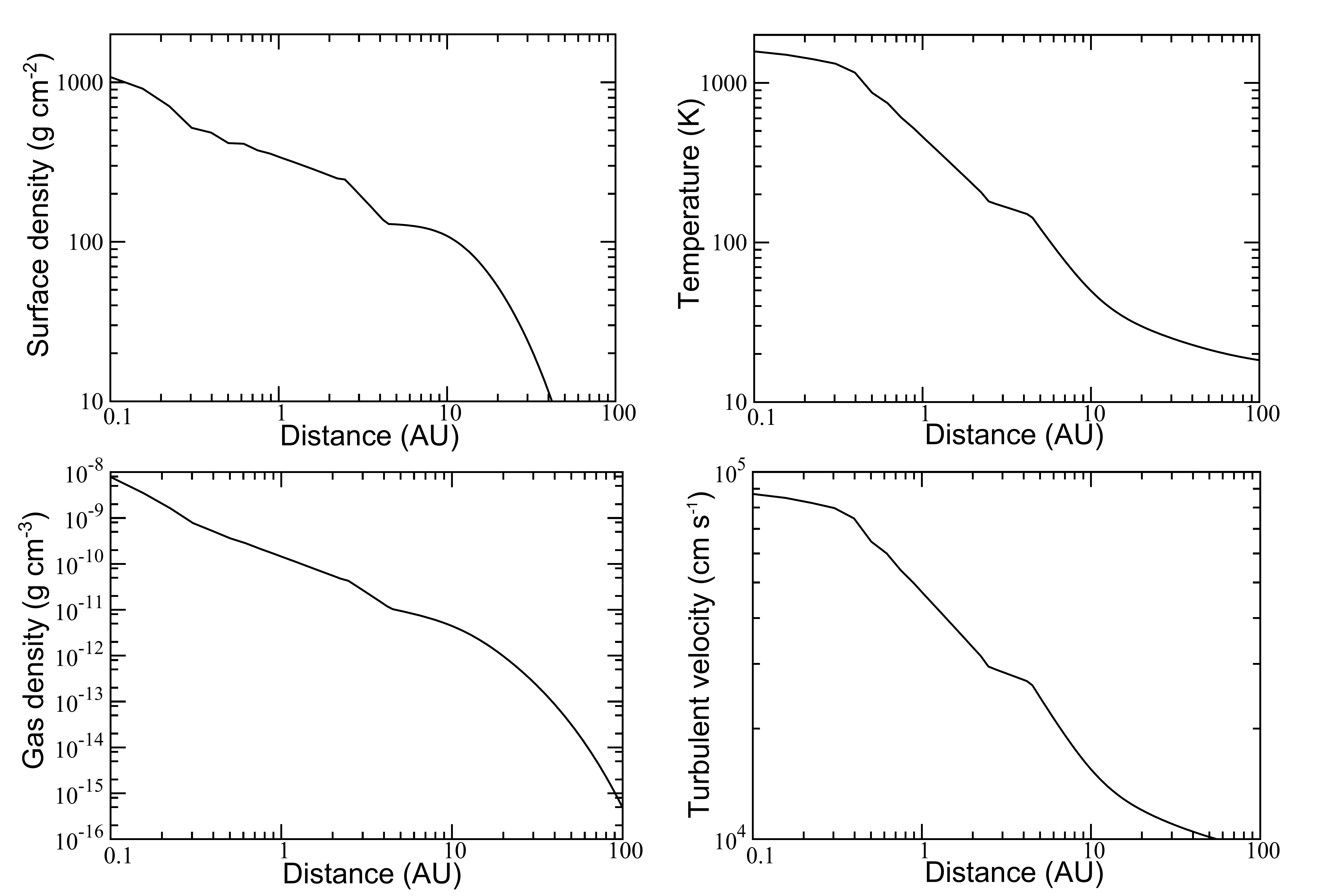}
\caption{Surface density (top left), midplane temperature (top right), gas density (bottom left) and turbulent velocity (bottom right) profiles of the used disk model.}
\label{fig:diskprop}
\end{center}
\end{figure*}

For the rest of this work, we will use the disk properties at 10$^{5}$ years. {The disk is presumed to be stationary since the planetesimal formation timescale is shorter than the disk evolution timescale.} 
The initial conditions of the disk model are those inferred by \cite{hg} (HG05) for the DM tau system: $M_{cloud}$ = 0.53 $M_\odot$, $\Omega_{cloud}$ = 23$\times$ $10^{-14}$ s$^{-1}$, $T_{cloud}$ = 16 K, $M_{0,star}$ = 0.01 $M_\odot$, $T_{star}$ = 4700 K but $\alpha$ = 0.01 (the value used by \cite{hughes} instead of $\alpha$ = 0.02 as in HG05). At $10^5$ years, this leads to a star--disk system with respective masses of 0.5 and 0.03 $M_{\odot}$. {We chose these initial conditions leading to a system less massive than our protosolar nebula for consistency. These parameters provided a best fit for the typical protoplanetary disk DM tau, and using a more massive disk will not change the qualitative results of this work since we are discussing abundances normalized with respect to solar value, so we preferred using consistent parameters instead of tweaking them.} The obtained disk properties are shown in Fig. \ref{fig:diskprop}. \\

In our model, {volatiles concentrations are presumed to be initially homogeneous} throughout the PSN with CO and N$_2$ supposed to be the main carriers of C and N \citep{prinn} and hence their abundances are set to the carbon and nitrogen solar abundances. Solids are assumed to be decimetric ``pebbles'' at their respective iceline\footnote{The region in protoplanetary disks where temperature becomes low enough to condense a volatile vapor.}. {{Centimetric} pebbles are observed in large quantities in disks \citep{wilner}. Interestingly, pebbles are found by models to have an optimal size for effective concentration in vortices and for accretion \citep{2012A&A...544A..32L}.} Inside the icelines there is only vapor. Since the sublimation temperatures for CO and N$_2$ are respectively 25 and 24 K \citep{fray09}, their icelines are located in our model at 28 and 32 AU. The CO iceline's position is comparable to that recently inferred at $\sim$ 30 AU in {the solar analogue} TW Hya \citep{CO1}. The exact sublimation temperature of these ices does not affect our scenario, it is only the difference between the two temperatures that is key to our results. The model then tracks the subsequent evolution of the system as a function of time and location.\\

A typical simulation starts with a decimetric pebble (N$_2$ or CO) near its corresponding iceline. This particle is large enough to decouple from gas. It will drift inward due to gas drag at the velocity determined by the transport module, and starts sublimating. The time needed for sublimation and the distance travelled before it happens are calculated by the sublimation module. These values are communicated to the vapor diffusion module through the source function. This module then evolves the vapor concentration inside the iceline. The vapor will diffuse outward along its concentration gradient due to existence of the iceline. The removed vapor will condense with time at the iceline into decimetric pebbles. These will get decoupled and start drifting inward repeating the cycle. The distribution of volatiles in our model is hence controlled by the balance of these two effects: the outward diffusion of the vapor and the ices inward migration followed by sublimation.

\begin{table*}
\begin{center}
\caption{Velocity, evolution time and distance for chosen solids sizes as found by our model.}
\begin{tabular}{lccccccc}

\noalign{\smallskip}
			&	\multicolumn{3}{c}{N$_2$}					&			\multicolumn{3}{c}{CO} 													\\
Size (cm) 		& Velocity (cm/s) 			& $\Delta t$ (years) 			& $\Delta R$ (AU) 		& Velocity (cm/s) 		& $\Delta t$ (years) 			& $\Delta R$  (AU) 		\\
\hline
10$^{-1}$ 		& - 					& 4.3$\times 10^{4}$		& - 					& - 				& 3.3$\times 10^{4}$		& - 					\\
1 			& -787.3 					& 1.3$\times 10^{4}$		& -22.5 				& -463.6 				& 2.0$\times 10^{4}$		& -19.0 				\\
10 			& -2647.0 					& 4.1$\times 10^{3}$		& -23.8 				& -2754.3 				& 3.7$\times 10^{3}$		& -20.8				\\
\hline

\end{tabular}\\
\end{center}
\label{tab:1}
{Notes.} Negative velocities mean inward drifts. $\Delta t$ is the time taken by 1 and 10 cm particles to drift from their starting positions until sublimation and for millimetric dust to grow into pebbles through vapor condensation (in this case $\Delta t \equiv t_{grow}^r$). $\Delta R$ is the distance travelled by inward drifting particles from their iceline to their sublimation location. The particles are placed initially on their iceline. Millimetric dust velocities and transport in RJ13 are dictated by turbulence and gas drag.
\end{table*}

\section{Results}
\subsection{Qualitative results discussion}


In our model the outward diffusion of vapor is shown to be faster than its replenishment inside the icelines by sublimating ices. This leads to depletion in vapors inside the icelines and a concentration of solids at the iceline positions. This result is of capital importance for this work.
Since CO is the major C-bearing volatile in the PSN, its iceline should be very rich in solids, explaining the origin of the high volumetric density of solids needed to form the planets. The high CO abundance in the building blocks implies that planets forming in this region should be very rich in carbon {in bulk}.\\

On the other hand the N$_2$ iceline is located slightly outward (4 AU) of the CO iceline. The proximity of the two icelines leads to a natural depletion in N$_2$ vapor at the CO iceline since the vapor diffusion depletes the area immediately inward of an iceline quicker than that further away. Therefore planets forming at the CO iceline should also be significantly depleted in nitrogen, compared to the solar N/H abundance.\\

Finally, coupling the D/H observations in Uranus and Neptune with our model where only a small fraction of the water present in the planets interiors is of nebular origin, and the rest originating from the transformation of CO into H$_2$O, leads to a higher D/H ratio for the proto-ices that formed the planets. The value found is compatible with internal structure models and the formation location of the planets in the same region as comets.\\

\subsection{Solids density and elemental abundances}
\begin{figure*}
\begin{center}
\includegraphics[scale=0.3]{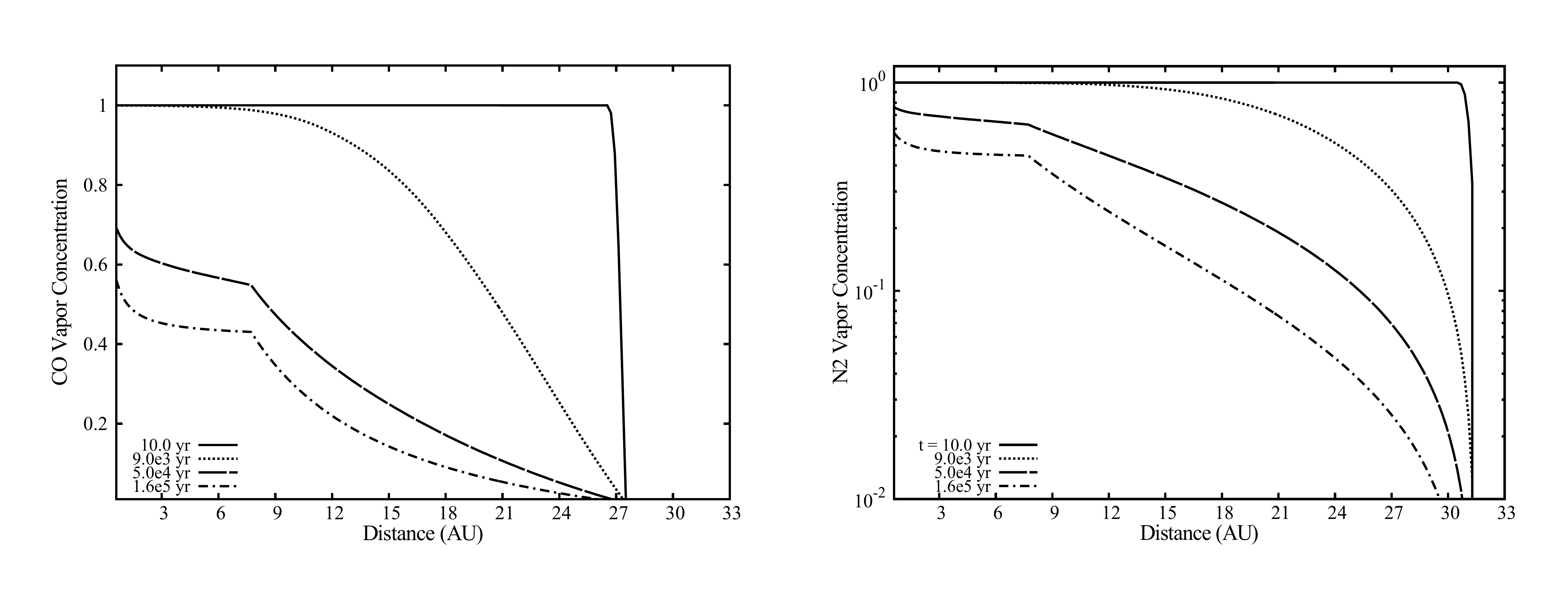}
\caption{{Vapors} concentrations of CO (left panel) and N$_2$ (right panel). The concentrations are normalized with respect to solar value. {Vapors} evolution is tracked inside their respective icelines as a function of time and distance to the star. In both cases there is a gradual location dependent depletion in the concentration due to gas diffusion being faster than replenishment through solid particles drift. N$_2$ is depleted by up to two orders of magnitude on the CO iceline.}
\label{fig:main}
\end{center}
\end{figure*}

\begin{figure}
\includegraphics[scale=0.32]{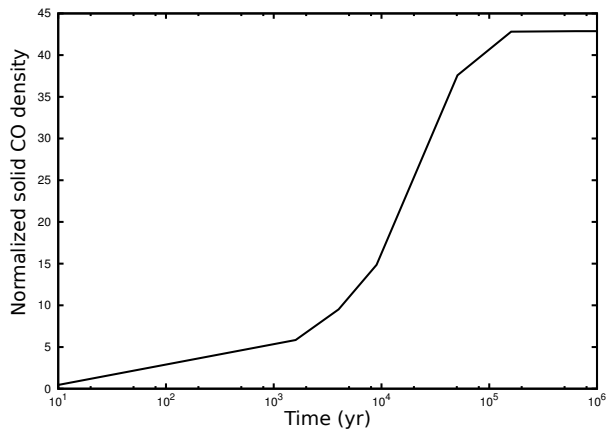}
\caption{The density of solid CO at its iceline, normalized with respect to solar value. Solid CO density increases as a function of time due to vapor diffusion from the inner nebula. In $2\times 10^{5}$ years, {the} density and chemical composition of this region becomes compatible with Uranus and Neptune.}
\label{fig:density}
\end{figure}

Figure \ref{fig:main} represents the evolution of CO and N$_2$ vapors inside their respective icelines. In {the final steady state at} 1.6$\times 10^{5}$ years, there is little vapor left inside these condensation fronts. All the missing vapor has been condensed into solids that concentrated at the icelines locations. {We note that the minor difference between this figure and its analogue in AD14 is due to a minor numerical correction from the last\footnote{See Erratum in preparation.}.} Figure \ref{fig:density} shows the evolution of solid CO normalized density as a function of time in the region near the iceline where all the CO ices concentrated. To quantify the solids surface density $\Sigma_s$ at the CO iceline we need first to calculate the length scale over which most of the diffused CO vapor condenses. We follow the prescription of \cite{stev88} in expressing this length scale as:
\begin{equation}
X_c=\delta\times ln(\sqrt{D\times t}/2\delta)
\end{equation}
where $t$ is the diffusion characteristic time, and $\delta$ is a parameter that measures the distance over which the saturation vapor pressure of the species changes significantly, and we use their $\delta = 0.1$ AU value. From our model we obtain $X_c \sim 0.5$ AU. This is the distance over which the solids will accumulate beyond the CO iceline.

\begin{figure*}
\begin{center}
\includegraphics[scale=0.5]{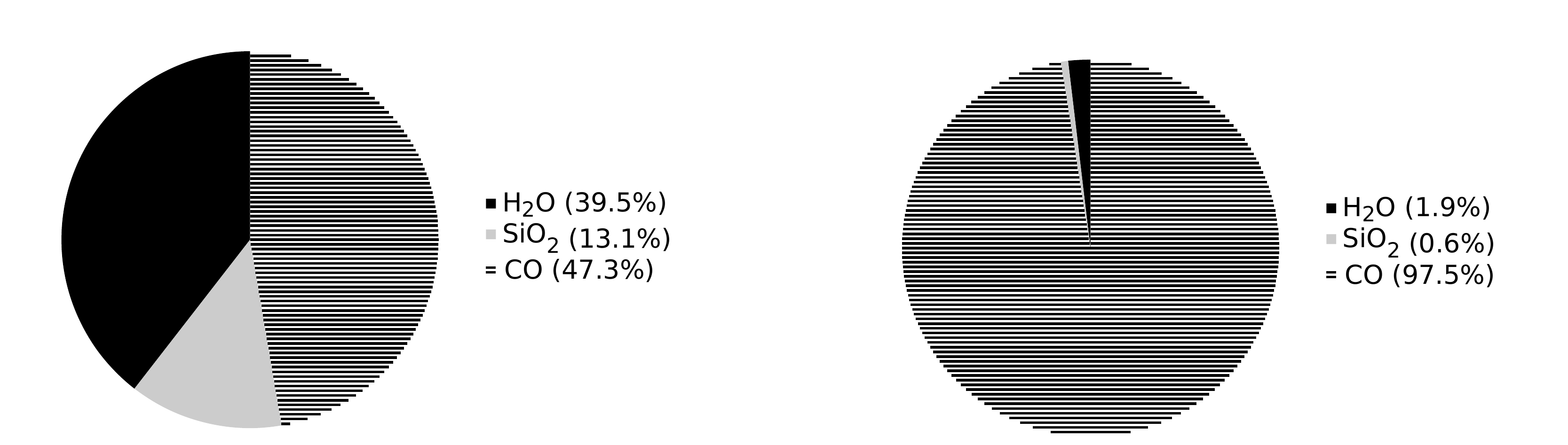}
\caption{{The mass fractions of solids at the CO iceline for $t$ = 0 yr (left) and $10^5$ yr (right). The left panel also describes the mass fractions between the CO and N$_2$ icelines (beyond the density peak) at any time. SiO$_2$ was included to represent silicate grains contribution, with solar Si abundance \citep{asplund}. At $t$ = $10^5$ yr, the mass distribution is almost completely dominated by CO ices.}}
\label{fig:pie}
\end{center}
\end{figure*}

The roughly 40 times solar CO enrichment in the planets formation zone, presented in Fig. \ref{fig:density}, is thus the integral of the CO vapor concentration removed from the entire inner region with 0.5 AU as integration step. 
Now we calculate the gas/solid $(G/S)$ mass ratio and the solids surface density $\Sigma_s$ in that region:
\begin{equation}
G/S=\frac{\Sigma_i \mu^g_i n^g_i}{\Sigma_j \mu^s_j n^s_j}
\end{equation}
where $\mu$ is the mean molecular weight of a species and $n$ is the molar abundance of gases (g) and solids (s). Taking into account CO and H$_2$O as solids, we obtain:
\begin{equation}
G/S = \frac{\mu_{H_2} \times n_{H_2}}{\mu_{CO}n_{CO}+ \mu_{H_2O}n_{H_2O}}
\end{equation}
We use $n_{H_2O}/n_{H_2}=7 \times 10^{-4}$ as found by \cite{cyr2} from chemical equilibrium calculations. We also chose $n_{CO}/n_{H_2O}=43\times 0.77$ which is the value observed in the inner region of AA Tauri's atmosphere \citep{carr} multiplied by the calculated enrichment factor. This value is higher than in comets, but lower than in the interstellar medium \citep{mumma}.
This gives finally $G/S = 3$ and $\Sigma_s = \frac{\Sigma_g}{G/S} = 9$ g cm$^{-2}$ (for $\Sigma_g = 27$ g cm$^{-2}$). {Figure \ref{fig:pie} shows the mass fractions of solids at the CO iceline before and after the CO vapor condenses.} The $\Sigma_s$ is more than one order of magnitude higher than the {initial value obtained for $G/S=72$.} This G/S value is more than enough to form the cores through gravitational collapses \citep{youdin2011}. \\
Moreover, \cite{dr2010} calculated that $6<\Sigma_s<11$ g cm$^{-2}$ is the best value to fit the formation timescales of Uranus and Neptune, {although the value we found is limited to a 0.5 AU wide zone.} {It should also be mentioned that the models of \citep{2009Icar..200..672D,dr2010} predict very large quantities of methane ices in the outer solar nebula, which is not observed in comets \citep{mumma}, but the authors stated that the CO/CH$_4$ ratio is not critical for their results. Moreover, in these models, Uranus and Neptune are formed with large amounts of ammonia, implying a large nitrogen abundance in their atmospheres, which is not observed (but not completely excluded).}

After core formation and the subsequent gas envelope accretion \citep{pol96}, the accreted CO will dissolve and transform into gaseous H$_2$O and CH$_4$ following
\begin{equation}
{\rm CO + 3H_2 \rightleftharpoons CH_4 + H_2O}
\end{equation}
resulting in the observed highly enriched atmospheric gaseous CH$_4$. {This nebular gas origin for the hydrogen in CH$_4$ is also consistent with the low CH$_3$D abundance measured in both planets \citep{feucht13,irwin2014}.}
Hence, the C/H and O/H ratios increases to more than 40 times the solar abundance. {The predicted C/H matches within the uncertainties the measured values of 34$^{+15}_{-11}$ and 48$^{+11}_{-13} \times$ solar for Uranus and Neptune respectively} (\cite{baines} using the solar abundances of \cite{asplund}).

Figure \ref{fig:main}b shows that at CO iceline location, N$_2$ vapor is depleted by {more than a factor of 50} with respect to solar value {after 2$ \times 10^5$ yr}. This implies that any planet forming in this region should be impoverished in nitrogen by factors similar to those inferred in Uranus and Neptune.\\

\subsection{The deuterium-to-hydrogen ratio}
To calculate the proto-ices D/H ratio in a manner consistent with Uranus and Neptune internal structures, previous works supposed that primordial water ice (and thus with cometary D/H value) {can represent up to $\sim$ 90\% of the planets mass \citep{helled11} (although it might be less due to rocks contribution).} This required the value of the proto-ices D/H to be $\sim 5\times 10^{-5}$ , a factor 6 lower (and can get up to an order of magnitude in some models) than the average cometary D/H value of $\sim 2-4\times 10^{-4}$ \citep{feucht13}. This led to speculations on the origin of their proto-ices and their interior structure. Using the observed planetary D/H for Uranus and Neptune, we perform the same calculations but assuming that most of the H$_2$O in the interior has CO as origin. 
Following \cite{feucht13} and assuming fully mixed envelopes, we can write:
\begin{equation}
(D/H)_{ices}=\frac{(D/H)_{planet}-x_{H_2}(D/H)_{gas}}{(1-x_{H_2})}
\end{equation}
where $(D/H)_{planet}$ is the value measured in Uranus and Neptune, $(D/H)_{ices}$ is the $D/H$ of the proto-ices accreted by these planets, $(D/H)_{gas}$ is the value for the PSN H$_2$ gas, supposed equal to the value in Jupiter's atmosphere. $x_{H_2}$ is the {molar} ratio of gas and ice defined as:
\begin{equation}
x_{H_2}=\frac{1}{1+\frac{(1-f_{H_2})}{(m_{H_2O}/m_{H_2})\times f_{H_2}}}
\end{equation}
where $m_{H_2O}$ and $m_{H_2}$ are the molar masses of H$_2$O and H$_2$ respectively. $f_{H_2}$ is the mass ratio of H$_2$ defined as: 
\begin{equation}
f_{H_2}=\frac{0.747M_{H_2+He}}{0.747M_{H_2+He}+M_{ice}}
\end{equation}
{where $M_{ice}$ is the total ices masses in the planets interiors (assumed to be H$_2$O), and $M_{H_2+He}$ is the total mass of Hydrogen and Helium.} 
Calculating $(D/H)_{ices}$ requires a prior knowledge of $x_{H_2}$ and therefore of the ice/gas ratio in the planets. In our model, the core accreted initially was dominated by CO ice. This CO then transforms into CH$_4$ and H$_2$O under the conditions obtained in the envelopes of Uranus and Neptune. For these two reasons, respectively, to calculate $(D/H)_{ices}$ inferred from our model, we use interior models of the planets from \cite{helled11} (H$_2$O model case 2) to determine $f_{H_2}$ and $x_{H_2}$, but we divide $M_{ice}$ ({corresponding to the ices mass fraction Z given in \cite{helled11}}) by 34 since $Z$ corresponds to the contributions of oxygen originating from both the protoplanetary CO and H$_2$O ices, while only the latter contributes to the D/H value. Assume $Z = Z_1 + Z_2$ where $Z_1$ and $Z_2$ are respectively the CO and H$_2$O ices contributions to $Z$. $Z_1 = 43 \times Z_{CO}^{solar} = 43 \times 0.77 \times Z_{H_2O}^{solar}$, since in our protoplanetary disk {CO is enriched 43 times over the solar value} {and $n_{CO}^{solar}/n_{H_2O}^{solar}=0.77$ as mentioned in the previous subsection.} Again we make the reasonable approximation that CO and H$_2$O are the dominant C and O bearing species. This leads to $Z_1 = 33 \times Z_{2}$ and $Z_2 = Z/(33+1)$. $Z_{2}$ is finally equal to $Z/34$.{ Hence we replace $Z$ in the original calculations by our $Z_2$ to calculate the D/H only for minor cometary water contribution to the global D/H ratio.} 

We then obtain for Uranus and Neptune: $f_{H_2}$ = 0.70 and 0.73 giving therefore $x_{H_2}$ = 0.954 and 0.960 respectively.
For $(D/H)_{planet}=4.2\times 10^{-5}$ and $(D/H)_{gas}=2.25\times 10^{-5}$ we can deduce $(D/H)_{ice}=4.4$ and $5.1$ $\times 10^{-4}$ for Uranus and Neptune respectively, values just slightly above the more enriched cometary D/H \citep{feucht13}. These calculations are a proof of concept, and can be improved using better interior models and initial water abundance estimates. {The methane contribution to the ices mass should also be taken into account for more precise calculations, instead of supposing that all ices are H$_2$O as in this work. We should also mention that an alternative explanation for the D/H problem might be found in the new model of \cite{2013Icar..226..256Y} where a non monotonic D/H gradient is inferred.}\\
{It should be noted that acquiring H$_2$O through CO transformation implies higher initial X/Z ratio. By reversing the problem we get $X'=0.230$, $Y'=0.024$ and $Z'=0.746$ initially during the planets formation. $X'/Y'$ is different from the solar value used by \cite{helled11}. The origin of this discrepancy is unclear and should be investigated.}

\section{Discussions}

\subsection{Consistency with dynamical models} 
The presence and initial positions of Uranus and Neptune are important for the Nice model that explains the orbital structure of the solar system \citep{n1,n2,n3}. An important component in the Nice model is the initial compact configuration of the giant planets, where Uranus and Neptune are at 15 and 12 AU respectively, much closer than the CO iceline. There are two possible way to reconcile this with our model. The first is that Uranus and Neptune probably migrated inward during the era of the gaseous disk from their formation location to the locations needed in the Nice model. Planets the mass of Uranus and Neptune are expected to undergo a type I migration due to their interaction with the disk gas. The characteristic timescale of type I migration is $10^4$ yr, and its usual direction is inward, although recent works showed that under some circumstances it can be outward \citep{guilet}. This migration will be halted when the ice giants enter a mean motion resonance (MMR) with Saturn and then each other. These are some of the initial conditions for the Nice 2 model \citep{morby07,morby072,levison}. Another possible solution is the migration of the CO and N$_2$ icelines themselves (due to the cooling of the gaseous disk over time), along with the accumulated matter, to these new locations prior to the formation of Uranus and Neptune. {On the other hand, the presence of Uranus and Neptune has minimal effects on the Grand tack scenario dealing with the solar system dynamics prior to disk dissipation\citep{walsh2011}. Finally we should mention the alternative model of \cite{2002AJ....123.2862T}, where Uranus and Neptune form between the orbits of Jupiter and Saturn. This scenario though is incapable of explaining the chemical compositions of the planets.}
 
\subsection{The model predictions}


The main prediction of the model we presented is the bulk O/H ratio in Uranus and Neptune. Since most of the oxygen is accreted in the form of CO, O/H should be $\sim$ C/H. Our prediction would imply primarily an external source of the observed stratospheric CO, {(steady micrometeorites influx or a {kilometric} sized cometary impact) a scenario consistent with recent observations \citep{lel10,lc13,cavalie2013,irwin2014}. An internal source necessitate a C/O ratio $\sim$ 0.1 (implying O/H $\geq 400\times$ solar value \citep{fl}), in contrast with our model where C/O $\sim$ 1. A future definitive observation of tropospheric CO on the other hand might imply C/O $\sim$ 0.1 according to the standard interpretation. This would apparently contradict the D/H measurement. All formation scenarios enriching O by such large fractions rely on accreting large cometary water ice quantities. This in turn leads to a high D/H for the planets, in contradiction with the observed values. The scenario of \cite{2004P&SS...52..623H} for example interpret the measured abundances of Uranus and Neptune's using a clathrates trapping model. Since almost 6 water molecules are needed for each gaseous volatiles to be trapped, this model predicted extremely high water abundance, inconsistent with the D/H measurement as mentioned above. Our model can also help constraining the Sulphur-to-Nitrogen ratio S/N, since other interpretations of the nitrogen depletion in Uranus and Neptune (nitrogen trapping in ammonia hydrosulfide clouds) necessitate a high S/N ratio where nitrogen has solar molar abundance and sulfur is supersolar \citep{gaut89,feg91}. On the other hand in our model, since nitrogen is fundamentally depleted (subsolar), then we anticipate a cometary abundance of Sulphur, which might be less or equal to the solar value \citep{jess,stardust}.\\}

\subsection{Effects on Jupiter and Saturn}
{Our model predicts a moderate depletion in gaseous CO and N$_2$ in the 3--6 AU Jupiter-Saturn formation region \citep{walsh2011}, but this does not necessarily contradict some formation models. For example, in \cite{2004ApJ...611..587L} carbon in Jupiter is supposed to be accreted from refractive carbonated materials, not gaseous CO. The scenario of \cite{gh} on the other hand does not explicitly discuss carbon enrichment in the giant planets, and the noble gases abundances are attributed to the disk atmosphere evaporation and dust settling. Finally, in the model of \cite{2005ApJ...626L..57A} where carbon is accreted through gas trapping in clathrates, a precise modeling of the planet formation is needed to understand the effect of this depletion.}\\

\subsection{Caveats}
\begin{figure}
\begin{center}
\includegraphics[scale=0.34]{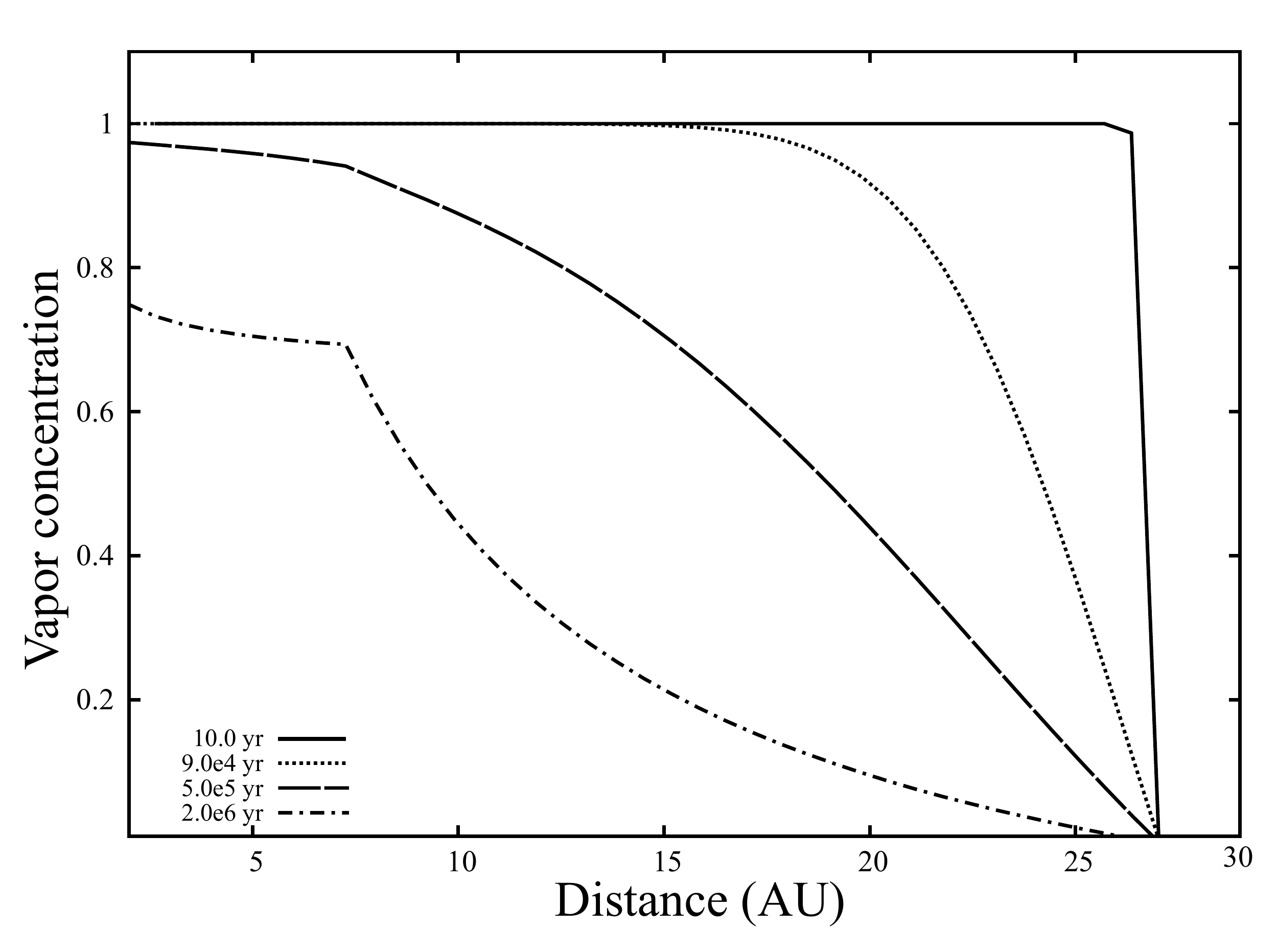}
\end{center}
\caption{Vapor concentration evolution in the case of a deadzone ($\alpha=10^{-4}$) for an iceline at 28 AU. We used $t_{grow}=10^4 \times \Omega^{-1}_K$ (RJ13). As for our nominal model, the vapor is still significantly depleted inside the iceline leading to solids enhancement beyond it. The evolution timescale is longer though, where steady state is reached only after $2\times 10^6$ yr.}
\label{fig:codz}
\end{figure}

{A first caveat is the assumed properties of the protoplanetary disk midplane. Our model assumed an outward advecting gas typically found in 2D viscous disks simulations and $\alpha \sim 0.01$. Recent MHD (Magneto-hydro-dynamics) simulations showed though that the midplane might be a MRI (Magneto-rotational-instability) inactive deadzone \citep{gammie,deadzone,COdead}, resulting in much weaker turbulence (and hence a smaller $\alpha$) and an almost static gas. {A lower $\alpha$ value decreases the diffusion rate ($D=3 \nu$) due to lower viscosities, but also significantly decrease the particles growth rate (RJ13)}. To simulate this aspect of dead zones, we run our model using $\alpha=10^{-4}$ (in all modules). Results are presented in Fig. \ref{fig:codz} showing the persistence of the solids enhancement effect even for weak turbulence.\\}

Another important component in our model is that both planets formed at the same location {in a narrow 0.5 AU sized region}, although not necessarily simultaneously. It might be possible for the growth of the first planet to be truncated by migration leaving enough solids behind to form the second. {Another possibility would be the first planet formation and migration before all the CO vapor is diffused throughout the CO iceline, allowing the second to form later from solids originating from the remaining vapor. The formation timescales are an obvious problem for this scenario, although the presence of a dead zone might significantly increase the CO ices concentration timescale. Using an evolving disk is an important step in this direction. A third possible solution would be if the diffused CO vapor condenses over a larger length scale $X_c$ (as those reported in RJ13), giving enough space for both planets to form simultaneously without interference. 
{Finally, advanced disk simulations that include the deadzone and gravitational instability effects \citep{martinsnow} show that the temperature gradient profile in protoplanetary disks might not be monotonous, allowing the possibility of having multiple icelines for the same specie at some stages of disk evolution, and thus hinting to another possible solution for this problem.}\\

{Finally, by interpreting the measured abundances of the ice giants as bulk compositions we assumed first that the reason for detecting the low NH$_3$ abundance is the intrinsically low bulk abundance. An alternative interpretation is that NH$_3$ condenses beyond the observable level. The second assumption is that both atmospheres are well mixed. This might not be the case for Uranus since its low heat flux indicates that it might not be fully convective, although this is unlikely for the more dynamic Neptune \citep{1993ARA&A..31..217L}.} \\


\section{Conclusions}
In this paper we showed how the formation on Uranus and Neptune on CO iceline resolves many issues related to these planets. The diffusive redistribution of vapor across the icelines increases the local solids density allowing the formation of these planets from carbon rich solids but nitrogen poor gas, and lead to planetary interiors consistent with recent D/H measurements.\\
Our scenario follows on from previous models \citep{stev88}, where Jupiter is formed on the H$_2$O iceline, a hypothesis to be firmly tested by \textit{Juno}. It expands this hypothesis to other planets and shows how this mechanism can solve certain long standing problems.
If it is true that most of the giant planets in our solar system were formed on icelines, it is difficult not to speculate that the same holds true for the formation of giant planets in general. 
In the last decade, hundreds of exoplanets has been discovered, with Neptune-mass bodies more abundant than Jovian-mass ones. 
In this work we gave a specific interpretation of a ``Neptune-like'' planet as one that formed on the CO iceline, {and is thus enriched with carbon with respect to the stellar value,} regardless of its mass. {This might allow the future generation telescopes to better classify planets with ambiguous masses and to disentangle mini-Neptunes and super-Earths.} It is possible that several mechanisms make Neptune-mass planets, and that those formed on the CO ice line and migrated inward are a subset of the total. {Only those to be found very enriched in carbon are Neptune-like according to this definition.} Testing this hypothesis will require compositional data on the atmosphere of such planets, of the sort we anticipate obtaining with the James Webb Space Telescope.
 
\acknowledgements
We thank T. Guillot and R. Hueso for having provided us with their accretion disk model. Special thanks go to T. Cavali\'{e} and E. Lellouch for useful discussions. We thank the anonymous referee for his useful comments that significantly improved the manuscript. M.A.-D was supported by a grant from the city of Besan\c{c}on. O.M. acknowledges support from CNES. JIL acknowledges support from the JWST program through a grant from NASA Goddard.


\begin{thebibliography}{}

\bibitem[Ali-Dib et al.(2014)]{mad} Ali-Dib, M., Mousis, O., Petit, J.-M., \& Lunine, J.~I.\ 2014, \apj, 785, 125

\bibitem[Alibert et al.(2005)]{2005ApJ...626L..57A} Alibert, Y., Mousis, O., Mordasini, C., \& Benz, W.\ 2005, \apjl, 626, L57 
 
\bibitem[Asplund et al.(2009)]{asplund} Asplund, M., Grevesse, N., Sauval, A.~J., \& Scott, P.\ 2009, \araa, 47, 481

\bibitem[Atreya et al.(1995)]{at95} Atreya, S.~K., Edgington, S.~G., Gautier, D., \& Owen, T.~C.\ 1995, Earth Moon and Planets, 67, 71 


\bibitem[Baines et al.(1995)]{baines} Baines, K.~H., Mickelson, M.~E., Larson, L.~E., \& Ferguson, D.~W.\ 1995, Icarus, 114, 328 

\bibitem[Carr \& Najita(2008)]{carr} Carr, J.~S., \& Najita, J.~R.\ 2008, Science, 319, 1504 

\bibitem[Cavali{\'e} et al.(2014)]{cavalie2013} Cavali{\'e}, T., Moreno, R., Lellouch, E., et al.\ 2014, \aap, 562, A33

\bibitem[Cyr et al.(1999)]{cyr2} Cyr, K.~E., Sharp, C.~M., \& Lunine, J.~I.\ 1999, \jgr, 104, 19003 

\bibitem[de Pater \& Richmond(1989)]{dep89} de Pater, I., \& Richmond, M.\ 1989, Icarus, 80, 1 

\bibitem[de Pater et al.(1989)]{dep892} de Pater, I., Romani, P.~N., \& Atreya, S.~K.\ 1989, Icarus, 82, 288 

\bibitem[Dodson-Robinson et al.(2009)]{2009Icar..200..672D} 
Dodson-Robinson, S.~E., Willacy, K., Bodenheimer, P., Turner, N.~J., 
\& Beichman, C.~A.\ 2009, Icarus, 200, 672 

\bibitem[Dodson-Robinson \& Bodenheimer(2010)]{dr2010} Dodson-Robinson, S.~E., \& Bodenheimer, P.\ 2010 Icarus, 207, 491 

\bibitem[Dzyurkevich et al.(2013)]{deadzone} Dzyurkevich, N., Turner, N.~J., Henning, T., \& Kley, W.\ 2013, \apj, 765, 114 

\bibitem[Fegley et al.(1991)]{feg91} Fegley, B., Jr., Gautier, D., Owen, T., \& Prinn, R.~G.\ 1991, Uranus, 147 

\bibitem[Feuchtgruber et al.(2013)]{feucht13} Feuchtgruber, H., Lellouch, E., Orton, G., et al.\ 2013, \aap, 551, A126 

\bibitem[Fray \& Schmitt(2009)]{fray09} Fray, N., \& Schmitt, B.\ 2009, \planss, 57, 2053 

\bibitem[Fletcher et al.(2009)]{fletcher2009} Fletcher, L.~N., Orton, G.~S., Teanby, N.~A., Irwin, P.~G.~J., \& Bjoraker, G.~L.\ 2009, Icarus, 199, 351 

\bibitem[Flynn et al.(2006)]{stardust} Flynn, G.~J., Bleuet, P., 
Borg, J., et al.\ 2006, Science, 314, 1731 


\bibitem[Gammie(1996)]{gammie} Gammie, C.~F.\ 1996, \apj, 457, 
355 

\bibitem[Gautier \& Owen(1989)]{gaut89} Gautier, D., \& Owen, T.\ 1989, Origin and Evolution of Planetary and Satellite Atmospheres, 487 

\bibitem[Giauque \& Clayton(1933)]{N2} Giauque \& Clayton.\ 1933, J. Am. Chem. Soc.\ 56, 4875 

\bibitem[Gomes et al.(2005)]{n1} Gomes, R., Levison, H.~F., Tsiganis, K., \& Morbidelli, A.\ 2005, \nat, 435, 466 



\bibitem[Guilet et al.(2013)]{guilet} Guilet, J., Baruteau, C., \& Papaloizou, J.~C.~B.\ 2013, \mnras, 430, 1764 

\bibitem[Guillot \& Hueso(2006)]{gh} Guillot, T., \& Hueso, R.\ 2006, \mnras, 367, L47 

\bibitem[Helled et al.(2011)]{helled11} Helled, R., Anderson, J.~D., Podolak, M., \& Schubert, G.\ 2011, \apj, 726, 15 

\bibitem[Helled 
\& Bodenheimer(2014)]{2014arXiv1404.5018H} Helled, R., \& Bodenheimer, P.\ 2014, arXiv:1404.5018 

\bibitem[Hersant et al.(2004)]{2004P&SS...52..623H} Hersant, F., Gautier, D., \& Lunine, J.~I.\ 2004, \planss, 52, 623 


\bibitem[Hueso \& Guillot(2005)]{hg} Hueso, R., \& Guillot, T.\ 2005, \aap, 442, 703 

\bibitem[Hughes \& Armitage(2010)]{hughes} Hughes, A.~L.~H., \& Armitage, P.~J.\ 2010, \apj, 719, 1633 

\bibitem[Irwin et al.(2014)]{irwin2014} Irwin, P.~G.~J., Lellouch, E., de Bergh, C., et al.\ 2014, Icarus, 227, 37 

\bibitem[Jessberger \& Kissel(1991)]{jess} Jessberger, E.~K., \& Kissel, J.\ 1991, IAU Colloq.~116: Comets in the post-Halley era, 167, 1075 

\bibitem[Lambrechts \& Johansen(2012)]{2012A&A...544A..32L} Lambrechts, M., \& Johansen, A.\ 2012, \aap, 544, A32 

\bibitem[Lellouch et al.(2010)]{lel10} Lellouch, E., Hartogh, P., Feuchtgruber, H., et al.\ 2010, \aap, 518, L152 

\bibitem[Levison et al.(2011)]{levison} Levison, H.~F., Morbidelli, A., Tsiganis, K., Nesvorn{\'y}, D., \& Gomes, R.\ 2011, \aj, 142, 152 


\bibitem[Lis et al.(2013)]{lis} Lis, D.~C., Biver, N., 
Bockel{\'e}e-Morvan, D., et al.\ 2013, \apjl, 774, L3 

\bibitem[Lodders(2004)]{2004ApJ...611..587L} Lodders, K.\ 2004, \apj, 611, 587 
\bibitem[Lodders \& Fegley(1994)]{fl} Lodders, K., \& Fegley, B., Jr.\ 1994, Icarus, 112, 368 

\bibitem[Lunine(1993)]{1993ARA&A..31..217L} Lunine, J.~I.\ 1993, \araa, 31, 217 

\bibitem[Luszcz-Cook \& de Pater(2013)]{lc13} Luszcz-Cook, S.~H., \& de Pater, I.\ 2013, Icarus, 222, 379 

\bibitem[Martin 
\& Livio(2012)]{martinsnow} Martin, R.~G., \& Livio, M.\ 2012, \mnras, 425, L6 

\bibitem[Martin \& Livio(2014)]{COdead} Martin, R.~G., \& Livio, M.\ 2014, \apjl, 783, L28 

\bibitem[Morbidelli et al.(2005)]{n2} Morbidelli, A., Levison, H.~F., Tsiganis, K., \& Gomes, R.\ 2005, \nat, 435, 462 

\bibitem[Morbidelli \& Crida(2007)]{morby07} Morbidelli, A., \& Crida, A.\ 2007, Icarus, 191, 158 

\bibitem[Morbidelli et al.(2007)]{morby072} Morbidelli, A., Tsiganis, K., Crida, A., Levison, H.~F., \& Gomes, R.\ 2007, \aj, 134, 1790 



\bibitem[Mumma \& Charnley(2011)]{mumma} Mumma, M.~J., \& Charnley, S.~B.\ 2011, \araa, 49, 471 


\bibitem[Pollack et al.(1996)]{pol96} Pollack, J.~B., Hubickyj, O., Bodenheimer, P., et al.\ 1996, Icarus, 124, 62 

\bibitem[Prinn(1993)]{prinn} Prinn, R.~G.\ 1993, Protostars and Planets III, 1005 

\bibitem[Qi et al.(2013)]{CO1} Qi, C., {\"O}berg, K.~I., Wilner, D.~J., et al.\ 2013, Science, 341, 630 

\bibitem[Ros \& Johansen(2013)]{ros} Ros, K., \& Johansen, A.\ 2013, \aap, 552, A137 

\bibitem[Stepinski \& Valageas(1996)]{step} Stepinski, T.~F., \& Valageas, P.\ 1996, \aap, 309, 301 

\bibitem[Stevenson \& Lunine(1988)]{stev88} Stevenson, D.~J., \& Lunine, J.~I.\ 1988, Icarus, 75, 146 

\bibitem[Supulver \& Lin(2000)]{sup} Supulver, K.~D., \& Lin, D.~N.~C.\ 2000, Icarus, 146, 525 

\bibitem[Thommes et al.(2002)]{2002AJ....123.2862T} Thommes, E.~W., Duncan, M.~J., \& Levison, H.~F.\ 2002, \aj, 123, 2862 

\bibitem[Tsiganis et al.(2005)]{n3} Tsiganis, K., Gomes, R., Morbidelli, A., \& Levison, H.~F.\ 2005, \nat, 435, 459 




\bibitem[Walsh et al.(2011)]{walsh2011} Walsh, K.~J., Morbidelli, A., Raymond, S.~N., O'Brien, D.~P., \& Mandell, A.~M.\ 2011, \nat, 475, 206 

\bibitem[Wilner et al.(2005)]{wilner} Wilner, D.~J., 
D'Alessio, P., Calvet, N., Claussen, M.~J., 
\& Hartmann, L.\ 2005, \apjl, 626, L109 


\bibitem[Wong et al.(2004)]{gal04} Wong, M.~H., Mahaffy, P.~R., Atreya, S.~K., Niemann, H.~B., \& Owen, T.~C.\ 2004, Icarus, 171, 153

\bibitem[Yang et al.(2013)]{2013Icar..226..256Y} Yang, L., Ciesla, F.~J., 
\& Alexander, C.~M.~O.~'.\ 2013, Icarus, 226, 256 

\bibitem[Youdin(2011)]{youdin2011} Youdin, A.~N.\ 2011, \apj, 731, 99 



 

\end{thebibliography}

\end{document}